\documentclass[journal]{IEEEtran}
\IEEEoverridecommandlockouts
\usepackage{cite}
\usepackage{amsmath,amssymb,amsfonts}
\usepackage[english]{babel}
\usepackage[utf8]{inputenc}
\usepackage[noend]{algpseudocode}
\usepackage{algorithm}
\usepackage{algpseudocode}
\usepackage{graphicx}
\usepackage{textcomp}
\usepackage{subcaption}
\usepackage{array}
\usepackage{tabularx}
\usepackage{xcolor}
\usepackage{bm}
\def\BibTeX{{\rm B\kern-.05em{\sc i\kern-.025em b}\kern-.08em
    T\kern-.1667em\lower.7ex\hbox{E}\kern-.125emX}}

\begin{document}

\title{Cooperative Dueling DQN–SAC Learning for Energy-Efficiency in Dynamic OWC Networks
}

\author{\IEEEauthorblockN{ Walter Zibusiso Ncube, Ahmad Adnan Qidan, Taisir El-Gorashi and Jaafar M. H. Elmirghani}

\thanks{Walter Zibusiso Ncube,
    Ahmad Adnan Qidan, Taisir El-Gorashi,
    and Jaafar M. H. Elmirghani, Department of Engineering, King’s College London, London, United Kingdom (emails: (walter.ncube, ahmad.qidan, taisir.elgorashi, jaafar.elmirghani)@kcl.ac.uk).}
}
\maketitle

\begin{abstract}
Growing wireless traffic is increasing pressure on the congested radio-frequency spectrum. Optical wireless communication (OWC) provides a complementary solution by using the abundant unlicensed optical spectrum. However, indoor OWC networks are dynamic: users move, enter or leave the network, and subscribe to different services. Poorly coordinated resource allocation can consequently waste subcarriers, require excessive transmission power and cause frequent AP reassignments. Energy efficiency (EE), defined as the total delivered data rate divided by the total network power consumption, therefore requires the serving AP, number of allocated subcarriers and transmission power to be jointly adapted while maintaining QoS. Optimising these in a dynamic time series, multi-service OWC environment produces a complex sequential EE optimisation problem. To address this problem, this work proposes Dual-Agent Resource Allocation using Deep Reinforcement Learning (DARA-DRL). DARA-DRL combines a branching duelling deep Q-network for association and subcarrier allocation with a conditional soft actor-critic agent for continuous power control. The agents are coupled through a common reward and a cooperative value update that evaluates each discrete allocation together with its corresponding power decision. Simulation results show that DARA-DRL remains within 5\% of the
optimal solution and compared with state-of-the-art benchmarks, it
improves EE by 28.8\% and
QoS satisfaction by 10.5\%, while reducing online decision
time by 14.5\%. Results demonstrate that agent specialisation simplifies mixed-action learning, and cooperation outperforms independently trained agents.
\end{abstract}


\section{Introduction}
\label{sec:introduction}

\IEEEPARstart{T}{he} rapid growth in wireless data traffic continues to put pressure on the limited radio-frequency (RF) spectrum. Optical wireless communication (OWC) provides a complementary solution by using the abundant optical spectrum to support high-data-rate indoor connectivity \cite{Ncube2023EE,Chowdhury2018Survey,Soltani2023LiFi}. Vertical-cavity surface-emitting lasers (VCSELs) are particularly attractive for this purpose because their wide modulation bandwidth and narrow beams support high transmission rates and concentrated optical coverage. Compared with light-emitting diode (LED)-based systems, VCSEL-based OWC can also achieve higher energy efficiency (EE) \cite{Ncube2023EE}. Previous studies examined how beam waist, micro-lenses and optical cell size affect achievable rate and EE \cite{Ncube2022EE}. These physical-layer improvements provide a basis for efficient transmission, but their network-level benefit also depends on how power and resources are shared among users.

EE is defined here as the total delivered data rate divided by the total network power consumption. Improving EE therefore requires more than reducing transmit power or increasing throughput independently \cite{Murdock2014,Ncube2024Cells}. An allocation must deliver useful data while accounting for transmitter and circuit consumption, and the overhead of changing the allocation. This balance is particularly important in dynamic indoor OWC networks, where users move, arrive and depart, and subscribe to services with different quality-of-service (QoS) requirements \cite{Ncube2025Predictive}. User movement changes optical channel gains, while arrivals and departures change AP loads and the resources available to existing users. Consequently, an allocation that provides a service efficiently in one interval may require more power or become insufficient in the next.

This work focuses on three coupled EE optimisation decisions: AP association, subcarrier allocation and transmit power. AP association determines the serving optical link and the users sharing an AP's resources. Subcarrier allocation determines the frequency resource available to each user, while power allocation affects the rate achieved over that resource. These decisions cannot generally be assessed independently \cite{Jiang2018,Liu2025JAPSRA}. A strong optical link may belong to an AP with little remaining capacity, while a less heavily loaded AP may require more power to serve the same user \cite{Jiang2018}. Allocating additional subcarriers can reduce the power needed to achieve a given rate, but leaves fewer resources for other users and introduces additional processing consumption. Moreover, subcarrier reuse by different APs couples the power decisions through inter-AP interference \cite{Jiang2018,Liu2025JAPSRA}.

The decisions are also linked across consecutive allocation intervals. Repeated AP reassignments and changes in subcarrier count require control signalling and resource reconfiguration. For example, a small improvement in channel gain near an AP coverage boundary may favour reassignment in the current interval, although the user may soon move back towards the previous AP \cite{Frometa2024Handover}. The resulting allocation changes can offset the initial EE benefit. Different service requirements further affect whether retaining or changing an allocation is appropriate: a configuration that remains adequate for one service may no longer meet another service's minimum rate \cite{Ncube2025Predictive,Liu2024JUGSAPA}. The controller must therefore balance current EE and QoS against the cost of modifying the previous allocation, rather than treating every interval as an unrelated problem.

Existing studies have addressed important aspects of OWC resource allocation and EE optimisation. In \cite{Jiang2018}, joint user association and power allocation in cell-free VLC networks was investigated, considering user fairness and load balancing through network-utility maximisation. The approach decomposed association and power allocation and solved the resulting subproblems iteratively. EE optimisation for rate-splitting indoor VLC was considered in \cite{Xing2022}. More recently, in \cite{Liu2025JAPSRA}, joint AP selection, sub-channel allocation and power allocation for EE optimisation in aggregated VLC--RF networks was investigated. The framework combined matching-based resource assignment with power optimisation, demonstrating the importance of coordinating the serving link and available resources. Related work considered user grouping, subcarrier allocation and power allocation to improve sum rate and user satisfaction in multicolour VLC--NOMA networks \cite{Liu2024JUGSAPA}.

These studies provide useful approaches for coordinating resources, but joint optimisation for a given network configuration does not by itself account for the consequences of successive allocation changes. In the dynamic problem considered here, the previous allocation affects the cost of the current decision, while changing user loads and service requirements affect its suitability over time. Repeated mathematical optimisation can also introduce a substantial online computational burden \cite{Jiang2018,Ncube2025Predictive}. Simple allocation rules reduce this burden, but a rule based only on the strongest current link or available capacity does not provide the best balance between EE, QoS and allocation stability. The challenge is therefore both to coordinate the resource decisions and to account for their effects over successive intervals.

Reinforcement learning (RL) provides a suitable basis for this sequential decision problem. Through interaction with the network environment, an RL policy learns from the reward associated with an allocation and the rewards obtained in subsequent intervals \cite{Mnih2015DQN,Haarnoja2018SAC}. It can therefore account for effects that are not captured by the immediate achieved rate alone. Unlike learning to reproduce a set of labelled allocation decisions, RL evaluates actions through their observed consequences \cite{Mnih2015DQN}. Training can be performed offline across different network conditions, after which the learned policy produces online decisions through network inference \cite{AlHammadi2022DQL}. Importantly, learning from successive rewards does not require predicted future positions or demands to be supplied as inputs; it uses the experience collected during training to assess the longer-term value of current actions \cite{Mnih2015DQN}.

Learning-based OWC resource allocation has been investigated using Q-learning for steerable laser-based systems \cite{Elgamal2021Steerable}, and deep Q-learning for joint power allocation and LED transmission-angle control in VLC networks \cite{AlHammadi2022DQL}. Cross Q-learning has also been applied to resource allocation in user-centric OWC networks \cite{Feng2025CrossQ}, while prediction-assisted RL has been considered for handover decisions in hybrid LiFi/Wi-Fi networks \cite{Frometa2024Handover}. These applications establish the relevance of learning to optical resource management. However, applying RL to the present problem requires a design that handles service-dependent QoS, discrete AP and subcarrier decisions, continuous power control and allocation-change costs within the same sequential framework.

The mixed action space makes this design important. A deep Q-network (DQN) supports discrete actions \cite{Mnih2015DQN}, but directly representing every combination of users' AP and subcarrier choices produces a large output space. Action branching provides a more compact representation of these discrete decisions \cite{Tavakoli2018Branching}. Using DQN for power control would nevertheless require dividing power into discrete levels, with a trade-off between allocation resolution and learning complexity. Soft actor--critic (SAC) supports continuous power decisions \cite{Haarnoja2018SAC}, but its continuous-action formulation does not directly enforce binary AP association or integer subcarrier counts. Rounding continuous outputs would also require further processing to respect the shared AP resource limits.

Hybrid-action methods such as P-DQN \cite{Xiong2018PDQN} and cooperative methods such as HAPPO \cite{Kuba2022HAPPO} offer alternatives, but their application still requires a suitable representation of the coupled allocation decisions. Assigning discrete allocation to a DQN and power control to SAC gives each learner a suitable action type, but does not specify how their updates should cooperate. The value of an AP--subcarrier decision depends on the power subsequently allocated and the resulting interference and QoS outcome. Accordingly, evaluating the discrete choice without accounting for the corresponding power decision can give incomplete feedback \cite{11370966}. A coordinated design should allow the power agent to observe the discrete allocation and the discrete agent to learn from the performance of the complete allocation.

This work develops Dual-Agent Resource Allocation using Deep Reinforcement Learning (DARA-DRL) for this purpose. A dynamic, multi-service VCSEL-based OWC network is modelled, and a long-term EE problem jointly considers AP association, DCO-OFDM subcarrier counts and continuous transmit power, including AP-reassignment and subcarrier-count reconfiguration costs. DARA-DRL uses a branching duelling DQN for discrete allocation, followed by a conditional SAC agent for power control. The agents share a system reward and replay transitions during centralised training \cite{Lowe2017MADDPG}. A cooperative value update uses the power agent's critics to evaluate the complete allocation, while the power agent conditions its output on the discrete decision. Feasibility processing enforces the association, subcarrier and AP power limits, and the reward penalises minimum-rate shortfalls. The resulting framework learns sequential allocation decisions from current measurements and previous allocations without requiring forecast information.

The main contributions of this work are summarised as follows:

\begin{itemize}

\item A long-term EE optimisation problem is formulated for a dynamic, multi-service VCSEL-based OWC network. It jointly determines AP association, DCO-OFDM subcarrier counts and continuous transmit power, while accounting for service requirements and the additional power associated with AP reassignments and subcarrier reconfiguration. The discrete decisions, non-convex rate-power relationships and dependence between successive allocations make repeated optimisation computationally demanding. The optimisation problem is reformulated as an MDP with an EE and QoS based reward. Its large state space and coupled discrete--continuous actions motivate an advanced DRL solution that learns allocation decisions from current measurements and previous allocations.

\item DARA-DRL is developed to tackle the optimisation problem. It uses two specialised agents: a branching duelling DQN for AP association and subcarrier-count selection, followed by a conditional SAC agent for continuous power control. Action branching avoids explicitly enumerating all users' joint discrete choices, while SAC avoids discretising transmit power. The agents are trained centrally using a common reward and shared replay transitions. A cooperative value update uses the power agent's critics to evaluate the complete allocation, allowing the discrete agent to account for the power consequences of its choices. In turn, the power agent conditions its decisions on the selected discrete allocation. During online operation, the trained agents act sequentially through forward inference, avoiding repeated iterative optimisation.

\item Constraint handling is integrated into both allocation stages. The discrete feasibility layer assigns each user to exactly one AP and accepts subcarriers whose total does not exceed each AP's capacity. A deterministic mapper assigns the selected number of subcarriers to each user, preventing users served by the same AP from sharing a subcarrier. Power projection ensures users are allocated power only from their assigned AP and that the total assigned power for users under each AP is within the respective AP's limits.

\item DARA-DRL is evaluated against six benchmarks: the Optimal Solution, PDP-OPA, reactive greedy, independent branching-DQN--SAC, P-DQN and HAPPO. Comparisons assess convergence, optimality gap, EE, minimum-rate satisfaction, AP reassignment and online decision time. Additional comparisons use two modified versions of DARA-DRL: one replaces the two specialised agents with a single hybrid agent, while the other removes cooperative updates. These comparisons assess the benefits of agent specialisation and cooperation, respectively.

\end{itemize}

The remainder of this paper is organised as follows. Section~\ref{sec:system_model} presents the system model. Section~III formulates the long-term EE optimisation problem. Section~IV describes DARA-DRL and its training procedure. Section~V presents the performance evaluation, and Section~VI concludes the paper.

\section{System Model}
\label{sec:system_model}

\subsection{Room and Network Setup}

\begin{figure}[t]
\centering 
\includegraphics[width=1\columnwidth,keepaspectratio]{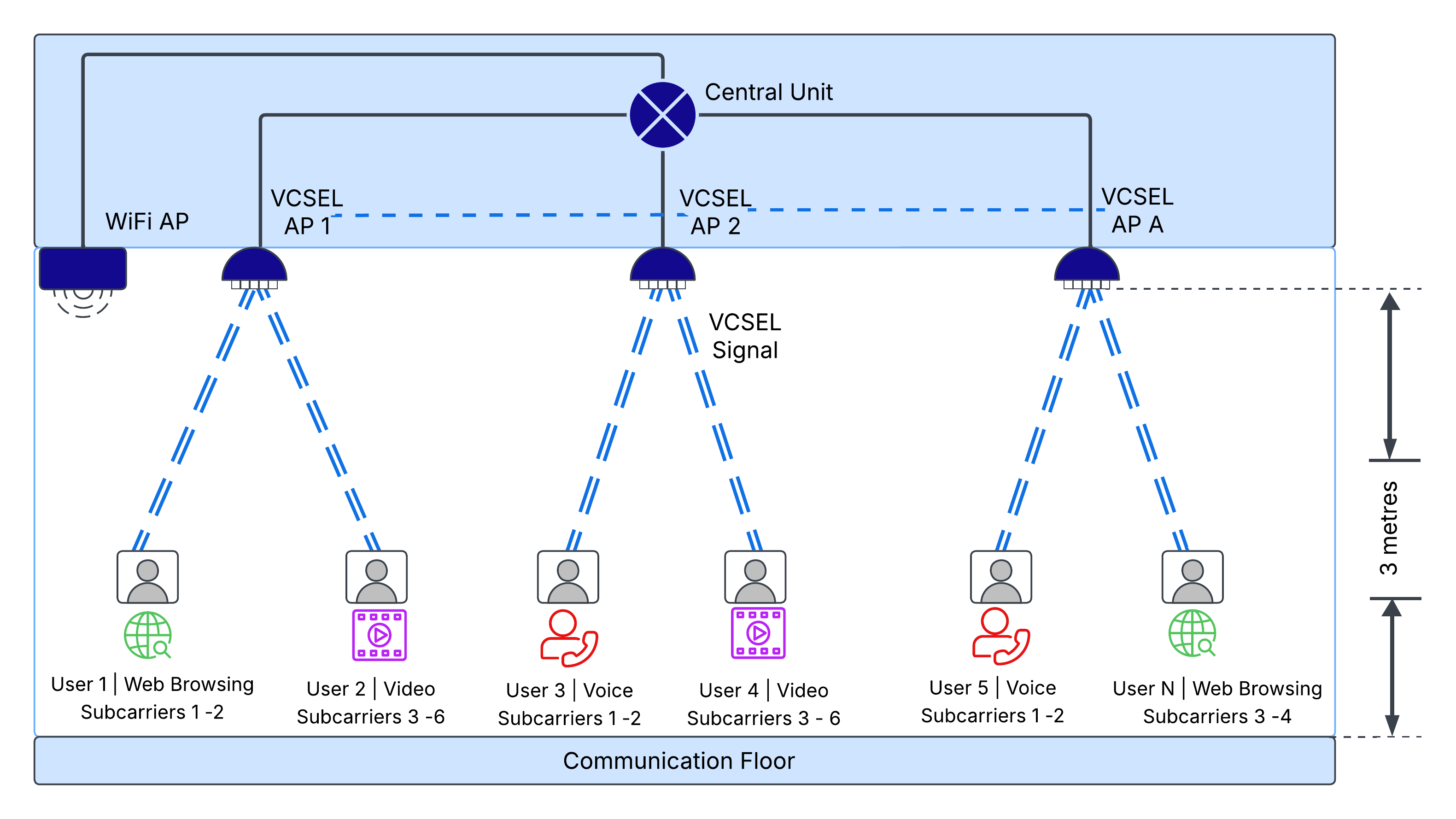} 
\caption{Indoor VCSEL-based OWC system.}
\label{systemmodel}
\end{figure}

Consider a downlink indoor OWC network with $\mathcal{A}=\{1,\ldots,A\}$  ceiling-mounted optical APs as in Figure~\ref{systemmodel}. Each AP contains an $L_{c}\times L_{c}$ micro-lens VCSEL array, giving $Q=L_{c}^{2}$ optical transmit elements. The VCSELs are used for high-speed data transmission and are not required for illumination. All APs are connected to a central unit (CU), which collects network information and coordinates the resource-allocation decisions. A Wi-Fi AP provides the uplink used for channel feedback, user requests and control signalling \cite{11673253,11164782}.

To capture user mobility and changing traffic conditions over time, network operation is divided into
$\mathcal{T}=\{T^{1},\ldots,T^{J}\}$ time slots, each of equal duration $\tau$, where $j\in\{1,\ldots,J\}$. Users may enter the room, leave the room or move between AP coverage
regions. Their position is treated as
constant during $\tau$ and are updated between consecutive slots.  The system is designed to support a number of users with time-varying demands denoted as $N$, \( n = \{1, \ldots, N\} \). The users active during slot $T^{j}$ are denoted as $\mathcal{N}^j$, \( n = \{1, \ldots, N^j\} \) and user $n\in\mathcal{N}^{j}$ is located at
$\mathbf{r}_{n}^{j}=[x_{n}^{j},y_{n}^{j},z_{n}^{j}]^{\mathsf T}$. Each user
is equipped with an angle-diversity receiver (ADR) containing
$N_{\mathrm{PD}}$ photodetector faces. Compound parabolic concentrators
are used to increase the collected optical power and provide a wider
reception angle \cite{Sarbazi2024Receiver}. The ADRs are assumed to remain upward-facing, with their photodetector
faces directed at different angles towards the ceiling to collect
signals from the ceiling-mounted APs. Users subscribe to \(K \), \( k = \{1,\ldots, K\} \), demand/service classes such as Video Streaming,  Voice Communication, and General Browsing. Class $k$ is characterised by
a minimum rate $C_{\min,k}$. 

At the beginning of slot $T^{j}$, the CU collects current user locations, service demands and channel estimates from the APs and active users. These measurements and the previous allocation form the observation supplied to the resource-allocation policy. The CU runs the policy to select each user's AP, subcarrier count and transmit power, and sends the decisions to the APs. At the next slot, new measurements are collected and the allocation is updated.

\subsection{Downlink System Model}

The downlink uses an $M$-point DCO-OFDM waveform. Hermitian symmetry leaves $L=\frac{M}{2}-1$ independent data-carrying subcarriers, represented by $\mathcal{L}=\{1,\ldots,L\}$ \cite{Tsonev2015,Sarbazi2024Receiver}. Let $x_{a,n,l}^{j}$ indicate whether subcarrier $l$ is scheduled for user $n$ by AP $a$. The number of subcarriers assigned to that AP-user link is therefore defined as:
\begin{equation}
    q_{a,n}^{j}=\sum_{l\in\mathcal{L}}x_{a,n,l}^{j},
    \qquad
    \mathcal{L}_{a,n}^{j}=\left\{l\in\mathcal{L}:x_{a,n,l}^{j}=1\right\}.
    \label{eq:subcarrier_count}
\end{equation}

The optical channel is assumed to be flat across the DCO-OFDM data
subcarriers. This means that, for a given AP-user link within one time
slot, all data subcarriers experience the same optical channel gain,
although this gain can differ between users and change over time.
This approximation removes frequency-dependent channel variations
from the model. Hence, to simplify resource allocation, the optimisation selects
the number of subcarriers $q_{a,n}^{j}\in\mathbb{Z}_{\geq 0}$
assigned to each user, where $\mathbb{Z}_{\geq 0}$ denotes the
non-negative integers, while their individual identities are
determined separately.

Different APs may reuse subcarriers, causing inter-AP interference despite the flat optical channel. A deterministic mapper converts the selected counts into subcarrier assignments, avoiding reuse within each AP and favouring lower-interference subcarriers \cite{6644236}. Once the subcarrier assignments are established, the continuous power decision $p_{a,n}^{j}$ specifies the total electrical modulation power
that AP $a$ allocates to user $n$ in slot $j$. This power is divided
equally across the user's assigned subcarriers as:
\begin{equation}
    p_{a,n,l}^{j}=\frac{p_{a,n}^{j}}{q_{a,n}^{j}},
    \qquad l\in\mathcal{L}_{a,n}^{j}.
    \label{eq:equal_power_split}
\end{equation}

This allows the system to focus on the number of allocated subcarriers and the user-level power budget. Per-subcarrier power loading can later be introduced by replacing $p_{a,n,l}^{j}$ with a separate continuous decision for each active subcarrier.

Two variables are used to measure changes in a user's resource
allocation between consecutive slots; the AP reassignment indicator $\chi_{n}^{j}$; 
\begin{equation}
    \chi_{n}^{j}
    =
    {\wp} \!\left\{a_{n}^{j}\neq a_{n}^{j-1}\right\},
    \label{eq:handover_indicator}
\end{equation}
and the change in the user's subcarrier allocation $\delta_{n}^{j}$;
\begin{equation}
    \delta_{n}^{j}
    =
    \left|q_{n}^{j}-q_{n}^{j-1}\right|,
    \qquad
    q_{n}^{j}
    =
    \sum_{a\in\mathcal{A}}
    s_{a,n}^{j}q_{a,n}^{j},
    \label{eq:subcarrier_reconfiguration}
\end{equation}
where $a_n^j$ is the serving AP of user $n$ in slot $j$, ${\wp}\{\cdot\}$ is an indicator function that means $\chi_n^j=1$ when the serving AP changes and $\chi_n^j=0$
when the user remains connected to the same AP. $s_{a,n}^{j}$ is equal to 1 when user $n$ is associated
with AP $a$ and 0 otherwise. $q_n^j$ is the total number of subcarriers assigned to user $n$ by its serving AP, $\delta_n^j$ is 0 when the number of subcarriers remains
unchanged and is positive when subcarriers are added or removed. These variables do not prevent resource changes. They measure how often and by how much the decision changes, allowing the long-term objective in Section~III to distinguish a necessary reconfiguration from a short-lived change that provides little EE benefit.

\subsection{Optical Channel Model}

Each optical AP employs a micro-lens VCSEL array\cite{Ncube2022EE,Ncube2024Cells}. The output of each VCSEL is represented by a Gaussian beam. For VCSEL $i$ of AP $a$, the optical intensity at radial distance $r$ and axial distance $z$ is \cite{Ncube2022EE,Ncube2024Cells,Soltani2022Safety}
\begin{equation}
    I_{a,i}(r,z)
    =
    \frac{2P_{a,i}^{\mathrm{opt}}}
    {\pi w_{a,i}^{2}(z)}
    \exp\!\left(
    -\frac{2r^{2}}{w_{a,i}^{2}(z)}
    \right),
    \label{eq:gaussian_intensity}
\end{equation}
where $P_{a,i}^{\mathrm{opt}}$ is the optical output power and $w_{a,i}(z)$ is the beam radius after the micro-lens. The calculation of $w_{a,i}(z)$ using Gaussian-beam propagation and the ABCD transformation is given in \cite{Ncube2022EE,Ncube2024Cells}. Similarly, the VCSEL operating region and eye-safety calculation follow \cite{Soltani2022Safety}. The resulting AP power limit is introduced in Section~III.

Let $d_{a,n}^{j}$ denote the Euclidean distance from AP $a$ to user $n$, $\phi_{a,i,n}^{j}$ the radiance angle from VCSEL $i$, and $\psi_{a,i,p,n}^{j}$ the incidence angle at photodetector face $p$. The axial and radial beam distances at the receiver are:
\begin{equation}
    z_{a,i,n}^{j}
    =
    d_{a,n}^{j}
    \cos\!\left(\phi_{a,i,n}^{j}\right),
    \qquad
    r_{a,i,n}^{j}
    =
    d_{a,n}^{j}
    \sin\!\left(\phi_{a,i,n}^{j}\right).
    \label{eq:beam_coordinates}
\end{equation}

Following the Gaussian-beam ADR model in \cite{Sarbazi2024Receiver,11673253}, the channel gain from VCSEL $i$ of AP $a$ to photodetector face $p$ of user $n$ is:
\begin{align}
    h_{a,i,p,n}^{j}
    ={}&
    \frac{
    2A_{\mathrm{PD}}
    T_f
    G_{\mathrm{CPC}}
    }
    {
    \pi w_{a,i}^{2}(z_{a,i,n}^{j})
    }\\
    \exp\!\left[
    -\frac{
    2(r_{a,i,n}^{j})^{2}
    }{
    w_{a,i}^{2}(z_{a,i,n}^{j})
    }
    \right]
    \nonumber
    &\times
    \cos\!\left(\psi_{a,i,p,n}^{j}\right)
    {\wp}\!\left\{
    0\leq\psi_{a,i,p,n}^{j}
    \leq\Theta_{\mathrm{FOV}}
    \right\},
    \label{eq:optical_channel_gain}
\end{align}
where $A_{\mathrm{PD}}$ is the photodetector active area, $T_f$ is the optical-filter gain, $G_{\mathrm{CPC}}$ is the concentrator gain and $\Theta_{\mathrm{FOV}}$ is the receiver field of view. ${\wp}\{\cdot\}$ is an indicator function that equals to one when the
incident beam is within the photodetector face's field of view and
zero otherwise, giving zero direct channel gain outside this range.

\subsection{SINR and Achievable User Rate}

Let $X_{l}$ denote the complex data symbol placed on subcarrier $l$. Hermitian symmetry is imposed as
\begin{equation}
    X_{M-l}=X_{l}^{*},\quad l=1,\ldots,L,
    \qquad X_{0}=X_{M/2}=0,
    \label{eq:hermitian_symmetry}
\end{equation}
so the $M$-point inverse fast Fourier transform produces a real waveform \cite{Tsonev2015,Sarbazi2024Receiver}. The normalised time-domain sample is:

\begin{equation}
x[b]=\frac{1}{\sqrt{M}}
\sum_{l=0}^{M-1}\alpha X_l
\exp\!\left(\mathrm{j}\frac{2\pi lb}{M}\right),
\quad b=0,\ldots,M-1,
\label{eq:dco_ofdm_signal}
\end{equation}
where $b$ indexes the time-domain samples within 1 OFDM symbol,  $\alpha=\sqrt{M/(M-2)}$ normalises the modulation waveform. A DC bias $I_{\mathrm{DC}}$ is then added to obtain the non-negative optical-drive signal for intensity modulation and direct detection (IM/DD).

Each AP constructs its waveform from the data symbols of its associated users, with the subcarrier assignments and allocated powers determining the symbols placed on each
subcarrier and their transmission amplitudes. At the receiver,
the photodetector converts the received optical signal into an
electrical signal. After DC-bias removal, FFT processing separates
the subcarriers. The received symbol of user $n$ from AP $a$
on assigned subcarrier $l$ is therefore represented as:

\begin{align}
Y_{a,n,l}^{j}
= {}& R_{\mathrm{PD}}h_{a,n}^{j}
\sqrt{p_{a,n,l}^{j}}X_{n,l}^{j} +
\nonumber\\
&\sum_{\substack{a'\in\mathcal{A}\\a'\neq a}}
\sum_{n'\in\mathcal{N}^{j}}
x_{a',n',l}^{j}R_{\mathrm{PD}}h_{a'\rightarrow n}^{j}
\sqrt{p_{a',n',l}^{j}}X_{n',l}^{j}
+V_{n,l}^{j},
\label{eq:received_subcarrier}
\end{align}
where $R_{\mathrm{PD}}$ is the photodetector responsivity, $h_{a,n}^{j}$ denotes the effective optical channel gain
from AP $a$ to user $n$ in slot $j$ after ADR combining and can be determined from \eqref{eq:dco_ofdm_signal}.  $h_{a'\rightarrow n}^{j}$ is the effective channel gain from interfering AP $a'$ to user $n$, and $V_{n,l}^{j}$ is receiver noise. The indicator $x_{a',n',l}^{j}$ is one only when AP $a'$ assigns subcarrier $l$ to user $n'$. The corresponding electrical SINR is \cite{Sarbazi2024Receiver,11673253}
\begin{equation}
    \gamma_{a,n,l}^{j}
    =\frac{R_{\mathrm{PD}}^{2}|h_{a,n}^{j}|^{2}p_{a,n,l}^{j}}
    {\sigma_{n,l}^{2}
    +\displaystyle\sum_{\substack{a'\in\mathcal{A}\\a'\neq a}}
    \sum_{n'\in\mathcal{N}^{j}}
    x_{a',n',l}^{j}R_{\mathrm{PD}}^{2}|h_{a'\rightarrow n}^{j}|^{2}
    p_{a',n',l}^{j}},
    \label{eq:sinr}
\end{equation}
where $\sigma_{n,l}^{2}$ contains the shot, thermal and background-noise components.

Using the lower bound for an IM/DD optical channel \cite{11673253,11164782}, the achievable rate of user $n$ in slot $j$ is:

\begin{equation}
C_n^j
=
\sum_{l\in\mathcal{L}_{a_n^j,n}^{j}}
\frac{2B}{M}
\log_2\!\left(
1+\frac{\exp(1)}{2\pi}
\gamma_{a_n^j,n,l}^{j}
\right),
\label{eq:user_rate}
\end{equation}
where $\mathcal{L}_{a_n^j,n}^{j}$ contains the assigned subcarriers for user $n$ from AP $a$ at time slot $j$,
and $B$ is the bandwidth.

\subsection{Network Power Consumption}

The network power model accounts for the consumption of the CU,
active APs, data transmission, together with the overhead
of changing the power and resource allocations \cite{Murdock2014,Ncube2025Predictive}. Expressing all contributions
as average powers over slot $T^j$, the total network power is:

\begin{align}
P_{\mathrm{tot}}^{j}
={}&P_{\mathrm{CU}}
+\sum_{a\in\mathcal{A}}z_a^j P_a^{\mathrm{cir}} +\frac{1}{\eta_{\mathrm{v}}}
\sum_{a\in\mathcal{A}}\sum_{n\in\mathcal{N}^{j}}
s_{a,n}^{j}p_{a,n}^{j} +
\nonumber\\
&P_{\mathrm{sc}}
\sum_{a\in\mathcal{A}}\sum_{n\in\mathcal{N}^{j}}
s_{a,n}^{j}q_{a,n}^{j}
+P_{\mathrm{ho}}\sum_{n\in\mathcal{N}^{j}}\chi_n^j
+P_{\mathrm{rec}}\sum_{n\in\mathcal{N}^{j}}\delta_n^j.
\label{eq:network_power}
\end{align}

In  (\ref{eq:network_power}), the fixed power $P_{\mathrm{CU}}$ covers the CU and
backhaul, while $P_a^{\mathrm{cir}}$ covers the bias and fixed
driver and circuit power of AP $a$. The indicator $z_a^j$
is 1 when the AP is active and 0 otherwise. $\eta_{\mathrm{v}}$ is the electrical-to-modulation-power efficiency, and $P_{\mathrm{sc}}$ is the incremental baseband and driver power for an active user-subcarrier assignment. The final 2 terms account for power and resource allocation changes. The indicator
$\chi_n^j$ records an AP reassignment, while $\delta_n^j$ measures
the absolute change in the user's assigned subcarrier count. $P_{\mathrm{ho}}$ and $P_{\mathrm{rec}}$ represent
the equivalent average power of one AP reassignment and a one-unit
subcarrier-count change, respectively. 

\section{Energy-Efficiency Optimisation Problem}
\label{sec:optimisation_problem}

Based on the system model presented in Section~\ref{sec:system_model}, the optimisation problem is formulated over discrete time slots. At the beginning of slot $j$, the CU uses the current network measurements and previous allocation to determine each active user's serving AP, subcarrier count and transmit power. It must determine the serving AP, the number of assigned subcarriers and the transmit power of every active user subscribing to different services. The aim is to  provide high EE at each time slot and, over time, whilst satisfying QoS requirements of different user classes and avoiding association or subcarrier changes whose short-term benefit does not justify their cost.

\subsection{Problem Formulation}

The network EE is evaluated over the $J$ slots of the operating
period. In each slot, resource allocation uses current network
measurements and previous allocations, without forecast information.
The EE over this period, $\eta_{\mathrm{period}}$, is
\begin{equation}
\eta_{\mathrm{period}}
=
\frac{\displaystyle
\sum_{j=1}^{J}\sum_{n\in\mathcal{N}^{j}}C_{n}^{j}}
{\displaystyle
\sum_{j=1}^{J}P_{\mathrm{tot}}^{j}}
\quad \text{bits/J}.
\label{eq:horizon_energy_efficiency}
\end{equation}
Since all slots have equal duration, this ratio equals the total
delivered data divided by the total consumed energy.

The EE optimisation problem is therefore written as:

\begin{equation}
\begin{aligned}
&\mathcal{P}_{1}:\quad
\underset{
\{\mathbf{s}^{j},\mathbf{q}^{j},\mathbf{p}^{j}\}_{j=1}^{J}
}{\operatorname{maximize}}
\quad \eta_{\mathrm{period}}
\\[1mm]
&\text{subject to:}
\\
&\mathrm{C1}:\quad
\sum_{a\in\mathcal{A}}s_{a,n}^{j}=1,&&\forall n\in\mathcal{N}^{j},\ \forall j,
\\
&\mathrm{C2}:\quad
1\leq q_{a,n}^{j}\leq L,&&q_{a,n}^{j}\in\mathbb{Z}_{\geq0},\ \forall a,n,j,
\\
&\mathrm{C3}:\quad
\sum_{n\in\mathcal{N}^{j}}q_{a,n}^{j}\leq L,&&\forall a,\ \forall j,
\\
&\mathrm{C4}:\quad
C_{n}^{j}\geq C_{\min,k_{n}^{j}},&&\forall n\in\mathcal{N}^{j},\ \forall j,
\\
&\mathrm{C5}:\quad
\sum_{n\in\mathcal{N}^{j}}
s_{a,n}^{j}p_{a,n}^{j}\leq P_{a}^{\max},&&\forall a,\ \forall j,
\\
&\mathrm{C6}:\quad
0\leq p_{a,n}^{j}\leq P_{a}^{\max}, && \forall a,n,j.
\end{aligned}
\label{eq:EE_Opt}
\end{equation}

C1 ensures that every active user is associated with only 1 AP and $s_{a,n}^{j}$ is equal to 1 when user $n$ is associated
with AP $a$ and 0 otherwise. C2 gives an associated user between 1 and $L$ data subcarriers and gives an unassociated AP-user pair 0 subcarriers. C3 limits the total number of user-subcarrier assignments at each AP to the $L$ available data subcarriers. C4 guarantees the minimum rate of the subscribed service class. C5 limits the total AP transmit power, where $P_{a}^{\max}$ is selected to satisfy the VCSEL linearity and eye-safety requirements. C6 ensures that an AP allocates power only to its associated users and restricts the association indicator to 0 or 1.

Problem $\mathcal{P}_{1}$ is a mixed-integer non-convex fractional problem. The association and subcarrier variables are discrete, while the transmit powers are continuous. For each user, the controller must choose a serving AP and decide how many subcarriers to assign, with at least one subcarrier required. Thus, even before power allocation is considered, a slot with $N$ users has high number of discrete combinations. The rate and power decisions are also coupled as in \eqref{eq:user_rate}. 

An optimal solution can be obtained by exhaustively enumerating the association and subcarrier decisions and solving the continuous power-allocation problem for every feasible candidate. Such methods provide an important benchmark, but their worst-case complexity grows exponentially and they must be re-run whenever the network changes.

Heuristic methods can simplify the optimisation problem
by using predefined rules to select AP associations, subcarrier
counts and transmit powers. The reactive greedy \cite{Jiang2021Joint} method follows
this approach, making allocation decisions from the current
network conditions with low computational cost. However, it does
not accommodate user movement or future demand, so an allocation
that appears favourable now may require further AP reassignments
or subcarrier changes shortly afterwards. These additional changes
can reduce EE over time, motivating a policy that considers both
current conditions and the future consequences of its decisions.

A predictive solution could solve $\mathcal{P}_{1}$ repeatedly
over the rolling window. Although it can anticipate mobility and
demand changes, the larger horizon increases the number of
optimisation variables. One such predictive approach is PDP-OPA \cite{Ncube2025Predictive},
which reduces the computational cost by predicting the next network
state and decomposing association and power allocation. Nevertheless, it does not jointly
optimise subcarrier counts and still performs an iterative power update.

The main limitation shared by these approaches is that a search process is required whenever the network changes. This motivates a learning-based solution. A DRL policy can be trained offline through repeated interaction with the dynamic environment and can then generate a decision through a forward pass during operation. More importantly, DRL treats the user association, subcarrier allocation and the power allocation as a sequence of related decisions and can learn whether an immediate improvement is worth its effect on later EE, QoS and AP reassignments.

\subsection{Markov Decision Process Reformulation}

The EE optimisation problem, $\mathcal{P}_{1}$, is reformulated as a Markov Decision Process (MDP) to enable its solution using RL. The MDP defines the network state, action, state transition, and reward, allowing a learning policy to make resource-allocation decisions based on the current network conditions and their effect on future network performance.

The MDP is defined by
\begin{equation}
\mathcal{M}
=
\left\langle
\mathcal{O},\mathcal{D},\mathcal{P},r,\beta
\right\rangle,
\label{eq:mdp}
\end{equation}
where $\mathcal{O}$ is the state space, $\mathcal{D}$ is the action space, $\mathcal{P}$ is the state-transition law, $r$ is the reward function, and $\beta\in(0,1)$ is the discount factor.

At the beginning of slot $t$, the policy observes the state $\mathbf{o}^{t}\in\mathcal{O}$ and selects a joint action $\mathbf{u}^{t}\in\mathcal{D}$. After the AP association, subcarrier, and power allocations are applied, the system produces a reward $r^{t}$ and transitions to the next state $\mathbf{o}^{t+1}$ according to $\mathcal{P}(\mathbf{o}^{t+1}\mid\mathbf{o}^{t},\mathbf{u}^{t})$.
The objective is therefore to learn a policy that maximises the long-term accumulated reward while satisfying the requirements of $\mathcal{P}_{1}$. Here, $t$ indexes RL environment interactions. Each interaction corresponds to one system time slot, so $t=j$ within each episode.

The state space $\mathcal{O}$ contains the network information
available for resource allocation. At slot $t$, the observed state $\mathbf{o}^{t}\in\mathcal{O}$ includes the measured network state
$\boldsymbol{\Omega}^{t}$. This include the active users, their service classes, channel conditions, AP loads, and previous AP-association, subcarrier, and power
decisions.

The action space $\mathcal{D}$ contains the joint AP-association, subcarrier-count, and transmit-power decisions. Based on the observed state, the policy selects
\begin{equation}
\mathbf{u}^{t}
=
\left[
s_{a,n}^{t},q_{a,n}^{t},p_{a,n}^{t}
,\middle|
a\in\mathcal{A},\ n\in\mathcal{N}^{t}
\right],
\label{eq:mdp_action}
\end{equation}
where $s_{a,n}^{t}$ indicates whether AP $a$ serves user $n$, $q_{a,n}^{t}$ denotes the number of subcarriers assigned to user $n$ by AP $a$, and $p_{a,n}^{t}$ denotes the corresponding transmit power. The AP-association and subcarrier-count variables are discrete, whereas the transmit-power variables are continuous. Thus, the MDP
has a mixed discrete-continuous action space. A feasible action must satisfy constraints C1-C6 of $\mathcal{P}_{1}$.

The reward evaluates the network performance resulting from the selected allocation. The effective EE in slot $t$ is defined as
\begin{equation}
\eta_{\mathrm{eff}}^{t}
=
\frac{\displaystyle
\sum_{n\in\mathcal{N}^{t}}C_{n}^{t}}
{\displaystyle P_{\mathrm{tot}}^{t}},
\label{eq:mdp_effective_ee}
\end{equation}
where $C_{n}^{t}$ is the achieved rate of user $n$ and $P_{\mathrm{tot}}^{t}$ is the total network power consumption.

A feasibility-aware reward is defined as
\begin{equation}
r^{t}=
\begin{cases}
    \displaystyle
    \frac{\eta_{\mathrm{eff}}^{t}}{\eta_{\mathrm{ref}}},
    C_{n}^{t}\geq C_{\min,k_{n}^{t}},
    \quad \forall n\in\mathcal{N}^{t},
    \\
    \displaystyle
    -\frac{1}{|\mathcal{N}^{t}|}
    \sum_{n\in\mathcal{N}^{t}}
    \left[
    \frac{C_{\min,k_{n}^{t}}-C_{n}^{t}}
    {C_{\min,k_{n}^{t}}}
    \right]^{+},
    &
    \text{otherwise}.
\end{cases}
\label{eq:mdp_reward}
\end{equation}
where $[x]^{+}=\max(x,0)$ and $\eta_{\mathrm{ref}}$ is a fixed reference EE used for reward normalisation. If all active users satisfy their minimum rate requirements, the reward is proportional to the achieved EE. If one or more users fail to satisfy their
requirements, the reward becomes negative according to the average relative rate shortfall. Thus, only users whose achieved rates are below their required rates contribute to the penalty. If no users are active, the reward is set to zero.

Feasible action construction enforces the association, subcarrier and AP power constraints. The minimum-rate requirement depends jointly on these decisions and inter-AP interference; it is therefore encouraged through the rate-shortfall penalty in
\eqref{eq:mdp_reward}, rather than guaranteed by action construction.

The learning objective is
\begin{equation}
\underset{\pi}{\operatorname{maximize}}\quad
\mathbb{E}_{\pi}
\left[
\sum_{\ell=0}^{\infty}
\beta^{\ell}r^{t+\ell}
\right],
\label{eq:mdp_policy_objective}
\end{equation}
where $\pi$ denotes the resource-allocation policy and $\beta$ is the discount factor that determines the influence of future rewards on the current decision. The expectation accounts for uncertainty in user mobility, service demand, and channel conditions. The resulting MDP provides a general formulation of $\mathcal{P}_{1}$ that can be solved using different DRL algorithms. The following subsections develop the
proposed two-stage approach for handling its mixed discrete-continuous action space.

\section{Dual-Agent Resource Allocation using Deep
Reinforcement Learning}
\label{sec:farad_rl}

In the preceding MDP, the controller observes the network state and selects each user's AP, subcarrier count and transmit power. Since the AP and subcarrier decisions are discrete, and the power is continous, the learning algorithm must therefore handle both decision types while maximising the accumulated reward defined in the MDP.

DQN can select the AP and subcarrier count, but cannot directly produce continuous power \cite{Mnih2015DQN}. Using DQN for all 3 decisions would require dividing power into discrete levels: fewer levels limit power resolution, while more levels increase
the action space \cite{Tavakoli2018Branching}. Whereas, SAC supports continuous power control \cite{Haarnoja2018SAC}, but applying its continuous outputs to AP and subcarrier decisions would require rounding and feasibility
repair. Methods that accommodate both decision types offer alternatives.
P-DQN combines discrete decisions and continuous parameters within
one hybrid-action agent \cite{Xiong2018PDQN}. However, the common
MDP reward reflects their combined outcome, making their individual
effects of the 2 decions types difficult to distinguish. HAPPO, on the other hand, enables cooperative multi-agent learning \cite{Kuba2022HAPPO}, but does not itself
provide separate DQN and SAC learning rules for the two decision types. Moreover, Independent branching-DQN-SAC \cite{Tavakoli2018Branching,Haarnoja2018SAC} provides this specialisation:
DQN selects the AP and subcarrier count, while SAC selects power.
However, their separate updates do not explicitly coordinate
how a discrete allocation is evaluated with its resulting power
decision.

The original MDP problem presented in \eqref{eq:mdp} jointly contains the
AP-association, subcarrier-count, and power decisions. To handle the
mixed discrete-continuous action space, the proposed DARA-DRL decomposes this
decision into a discrete resource allocation stage followed by a
conditional continuous power allocation stage. A Discrete Resource Agent (DRA) determines the
AP association and the number of subcarriers assigned to each active
user, while a Continuous Power
Agent (CPA) subsequently determines the corresponding
transmit powers.  The DRA first selects each active user's
AP and subcarrier count. A deterministic mapper then assigns
the corresponding subcarrier identities. The DRA uses a branching duelling DQN with Double-DQN-style
separation between action selection and target evaluation
\cite{Mnih2015DQN,VanHasselt2016DoubleDQN,Wang2016Dueling,Tavakoli2018Branching}. The CPA uses the same current observation, together with the discrete allocation and resulting subcarrier information, to select transmit power. The CPA uses conditional SAC with 2 centralised critics \cite{Haarnoja2018SAC}. By using DQN for discrete allocation and SAC for continuous power
control, the proposed DARA-DRL matches each algorithm to its decision type. Both agents use the same transition, system reward and shared
replay buffer \cite{Mnih2015DQN,Haarnoja2018SAC}. During the DRA
update, the CPA critics evaluate the complete discrete-continuous
action, allowing the DRA to account for the power allocation
associated with its decision. In turn, the CPA learns power
allocations conditioned on the DRA's selected configuration. By considering rewards over successive slots, both agents learn to balance immediate EE against later QoS shortfalls and resource-reconfiguration costs.

\subsection{Discrete Resource Allocation Agent}

For each active user $n$, the DRA selects one
combined discrete action
\begin{equation}
    d_{n}^{t}
    =
    (a_{n}^{t},q_{n}^{t})
    \in\mathcal{D}_{n}
    =
    \mathcal{A}\times\{1,\ldots,L\},
    \label{eq:discrete_user_action}
\end{equation}
where $a_{n}^{t}$ denotes the selected AP and $q_{n}^{t}$ denotes the requested number of subcarriers. No resources are assigned to an inactive user. For an active user, the selected AP determines the association indicator $s_{a,n}^{t}$, while the selected subcarrier count determines $q_{a,n}^{t}
    =
    s_{a,n}^{t}q_{n}^{t}.$
Consequently, the DRA output defines the discrete part of the original
MDP action, while the continuous power variables $p_{a,n}^{t}$ are
determined subsequently by the CPA.

The DRA uses a shared state encoder followed by a separate action
branch for each supported user, following the branching architecture
in \cite{Tavakoli2018Branching}. The shared encoder extracts common
network information from the MDP state $\mathbf{o}^{t}$, while each
user branch produces action values for the corresponding active user.
For branch $n$, the $A L$ possible AP-subcarrier-count combinations
are evaluated, providing a ranking of the discrete choices available
to that user. The dueling representation for branch $n$ is given by
\begin{align}
Q_{\mathrm{D},n}(\mathbf{o},d_n;\theta_{\mathrm{D}})
={}&
V(\mathbf{o};\theta_V)
+A_n(\mathbf{o},d_n;\theta_{A_n})
\nonumber\\
&-\frac{1}{|\mathcal{D}_n|}
\sum_{\bar d_n\in\mathcal{D}_n}
A_n(\mathbf{o},\bar d_n;\theta_{A_n}),
\label{eq:dra_dueling_q}
\end{align}
where $V(\cdot)$ estimates the value of the network state and
$A_n(\cdot)$ measures the relative value of the possible
AP-subcarrier-count choices for user $n$
\cite{Wang2016Dueling}. During training, the DRA uses
$\epsilon$-greedy exploration \cite{Mnih2015DQN}, whereas during
online execution the learned branch values are used greedily.

The branch outputs are converted into a feasible joint discrete
allocation by a deterministic feasibility layer. This step is
necessary because independently selecting the highest-valued action
for every user may result in an allocation that exceeds the
available subcarriers of an AP. Initially, each AP $a$ has $L$
available subcarriers. The feasibility layer considers the candidate
AP-subcarrier-count choices produced by the active-user branches and
repeatedly accepts the highest-valued feasible candidate $(n,a,q)$.
A candidate is accepted only if its requested subcarrier count fits
within the remaining resources of AP $a$ and sufficient resources
remain to provide at least one subcarrier to every unassigned active
user. Candidates that violate these conditions are skipped and the
next feasible candidate is considered. The procedure continues until
every active user has been assigned an AP and a positive subcarrier
count. The resulting joint discrete allocation is expressed as
\begin{equation}
    \mathbf{d}^{t}
    =
    \mathcal{F}_{\mathrm{D}}
    \left(
    \left\{
    Q_{\mathrm{D},n}
    (\mathbf{o}^{t},\cdot;\theta_{\mathrm{D}})
    \right\}_{n=1}^{N_{\max}},
    \mathbf{m}^{t}
    \right),
    \label{eq:discrete_feasibility_layer}
\end{equation}
where $N_{\max}$ is the maximum number of simultaneously supported
users, $\mathbf{m}^{t}$ identifies the active-user rows, and
$\mathcal{F}_{\mathrm{D}}$ denotes the deterministic feasibility
function. The feasibility layer therefore converts the individual
branch decisions into a joint allocation satisfying the discrete
resource constraints C1-C3. Provided that all APs are considered as association choices
and $|\mathcal{N}^{t}|\leq AL$, the feasibility layer assigns
each active user to one AP with at least one subcarrier.
During allocation, sufficient subcarriers are reserved for
the remaining users without exceeding any AP's capacity.
The resulting allocation satisfies C1--C3.

After the discrete allocation has been determined, the
deterministic mapper constructs the subcarrier sets
$\mathcal{L}_{a,n}^{t}$ without changing the selected AP
or subcarrier count. For each user, it selects the required
number of subcarriers from those not already assigned by
the serving AP. Within an AP, each
subcarrier is assigned to at most one user, while different
APs may reuse the same subcarrier.

\subsection{Continuous Power Agent}

The CPA determines the continuous transmit-power allocation
conditioned on the feasible discrete allocation produced by the DRA.
Given the network state $\mathbf{o}^{t}$ and the discrete allocation
$\mathbf{d}^{t}$, the CPA uses a Soft Actor-Critic (SAC) policy to
generate a requested power vector:
\begin{equation}
    \widetilde{\mathbf{p}}^{t}
    \sim
    \pi_{\phi}
    \left(
    \cdot\mid\mathbf{o}^{t},\mathbf{d}^{t}
    \right),
    \label{eq:conditional_sac_policy}
\end{equation}
where $\phi$ denotes the trainable actor parameters. Conditioning the
power policy on $\mathbf{d}^{t}$ ensures that power is allocated only
after the AP association and subcarrier-count decisions have been
established. Thus, the CPA does not alter the discrete allocation and
instead determines the continuous component of the original MDP
action.

During training, the actor produces the mean and standard
deviation of a Gaussian distribution for each active user.
A sampled value is passed through a sigmoid function and
multiplied by the maximum power of the selected AP. This
gives the requested user power $\widetilde{p}_n^t$ within
the range from zero to $P_a^{\max}$. Outputs corresponding
to inactive users are set to zero. Since the combined
requests may exceed an AP's power budget, the requested
powers are scaled when necessary. For user $n$ associated
with AP $a$, the executed transmit power is obtained as
\begin{align}
p_{a,n}^{t}
&=
s_{a,n}^{t}
\widetilde{p}_{n}^{t}
\rho_{a}^{t},
\label{eq:power_projection}\\
\rho_{a}^{t}
&=
\begin{cases}
1,
& \displaystyle
\sum_{m\in\mathcal{N}^{t}}
s_{a,m}^{t}\widetilde{p}_{m}^{t}
\leq P_a^{\max},\\[3mm]
\dfrac{P_a^{\max}}
{\displaystyle
\sum_{m\in\mathcal{N}^{t}}
s_{a,m}^{t}\widetilde{p}_{m}^{t}},
& \text{otherwise},
\end{cases}
\label{eq:ap_power_scaling}
\end{align}
where $P_a^{\max}$ is the maximum transmit power of AP $a$.
The complete power projection is denoted by
\begin{equation}
\mathbf{p}^{t}
=
\mathcal{F}_{\mathrm{P}}
\left(
\widetilde{\mathbf{p}}^{t},
\mathbf{d}^{t}
\right).
\label{eq:power_feasibility_layer}
\end{equation}
The projection sets the power of unassociated users to zero and
proportionally scales the requested powers only when the total
requested power at an AP exceeds its available power budget. Hence,
the executed allocation satisfies the AP power constraints C5--C6
for every action. If the requested powers already satisfy the AP
budgets, $\rho_a^t=1$ and the requested powers are applied without
modification.

The power projection does not explicitly enforce the QoS constraint
C4. Satisfying the minimum-rate requirement depends jointly on the
selected AP, the number of allocated subcarriers, transmit power,
channel conditions, and inter-AP interference. Enforcing C4 through
a simple power projection could therefore modify the learned resource
allocation or conflict with the AP power budgets. Instead, QoS is
incorporated into the common reward in \eqref{eq:mdp_reward}, where
rate-shortfall penalties are applied whenever one or more active
users fail to satisfy their service-class requirements. The CPA can
therefore learn power allocations that balance energy efficiency
and QoS while respecting the hard power constraints.

The CPA employs two critics,
$Q_{\mathrm{C},1}(\mathbf{o},\mathbf{d},\mathbf{p};\psi_{1})$ and
$Q_{\mathrm{C},2}(\mathbf{o},\mathbf{d},\mathbf{p};\psi_{2})$, which
take the network state, the feasible DRA allocation, and the executed
power allocation as inputs. Consequently, each critic evaluates the
continuous power decision in the context of the complete
discrete-continuous allocation rather than evaluating power
independently of the selected AP and subcarrier configuration. This
centralized action information is available during cooperative
training and allows the CPA to capture the dependence of the global
reward on both resource-allocation stages \cite{Lowe2017MADDPG}.

Following SAC \cite{Haarnoja2018SAC}, the minimum of the two critic
estimates is used when constructing the learning target to reduce
overestimation caused by function approximation. The entropy term
encourages sufficient exploration during training, while the entropy
temperature $\alpha$ is adjusted towards the target entropy
$\mathcal{H}_{\mathrm{tar}}$. The target actor and target critics are updated gradually
using the corresponding current-network parameters
to provide stable learning targets. Since the CPA is conditioned on
the DRA output, the resulting policy learns the power allocation that
best complements the selected AP and subcarrier counts
with respect to the common system reward.

The resulting executed action of the two-stage allocation process is
therefore $\mathbf{u}^{t} =\left(\mathbf{d}^{t},\mathbf{p}^{t}\right),$ where $\mathbf{d}^{t}$ is the feasible DRA AP-subcarrier allocation and $\mathbf{p}^{t}$ is the feasible transmit
power allocation generated by the CPA. Together, these decisions
define the AP association, subcarrier allocation, and transmit powers
of the original MDP action in \eqref{eq:mdp_action}.

\subsection{Cooperative Learning Between the Agents}

The DRA and CPA are trained cooperatively using the same network
interaction and common reward. At each step, the joint transition is
defined as:
\begin{equation}
\left(
\mathbf{o}^{t},\mathbf{d}^{t},\mathbf{p}^{t},r^{t},
\mathbf{o}^{t+1},\zeta^{t}
\right),
\label{eq:joint_transition}
\end{equation}
where $\zeta^{t}$ is the terminal indicator, with $\zeta^{t}=1$ when
the episode terminates and $\zeta^{t}=0$ otherwise. The transition is
stored in a shared replay buffer $\mathcal{B}$, following the
off-policy learning procedure used by DQN and SAC
\cite{Mnih2015DQN,Haarnoja2018SAC}. Since both agents sample the same
transitions, their updates are based on the same network state,
allocation outcome, and global reward.

For a sampled next state $\mathbf{o}'$, the online DRA first generates
the next discrete allocation using its branch values and the
feasibility layer:
\begin{equation}
\mathbf{d}^{+}
=
\mathcal{F}_{\mathrm{D}}
\left(
\left\{
Q_{\mathrm{D},n}
(\mathbf{o}',\cdot;\theta_{\mathrm{D}})
\right\}_{n=1}^{N_{\max}},
\mathbf{m}'
\right).
\label{eq:next_discrete_action}
\end{equation}
The selected discrete allocation is then provided to the
CPA's target actor, which generates the next requested
power vector:
\begin{equation}
\widetilde{\mathbf{p}}^{+}
\sim
\pi_{\bar{\phi}}
\left(
\cdot\mid\mathbf{o}',\mathbf{d}^{+}
\right),
\end{equation}
where $\bar{\phi}$ denotes the parameters of the slowly
updated target actor
and the power projection produces the feasible power allocation:
\begin{equation}
\mathbf{p}^{+}
=
\mathcal{F}_{\mathrm{P}}
\left(
\widetilde{\mathbf{p}}^{+},
\mathbf{d}^{+}
\right).
\label{eq:next_projected_power}
\end{equation}
Thus, the next-state value of the DRA decision is evaluated together
with the CPA's continuous power allocation.

The CPA is trained using the twin-critic SAC framework. The critic losses use the executed allocations stored in
the replay buffer, whereas the actor loss evaluates newly
sampled powers after projection. The entropy term is
calculated from the requested powers before projection,
with the sigmoid transformation and power scaling
accounted for in the policy log-probability. Its critics
$Q_{\mathrm{C},1}(\mathbf{o},\mathbf{d},\mathbf{p};\psi_{1})$ and
$Q_{\mathrm{C},2}(\mathbf{o},\mathbf{d},\mathbf{p};\psi_{2})$
evaluate the power allocation conditioned on the complete discrete
allocation. The minimum of the two critic estimates is used to reduce
overestimation, while the entropy term in SAC maintains exploration
during training \cite{Haarnoja2018SAC}. The target actor and target critics are updated gradually
using the corresponding current-network parameters, and the entropy temperature $\alpha$ is adjusted towards
the target entropy $\mathcal{H}_{\mathrm{tar}}$.

The main cooperation mechanism is introduced in the DRA update.
Instead of evaluating the next discrete allocation using only its
branch values, the target CPA critics evaluate the complete
discrete-continuous allocation. The cooperative DRA target is:
\begin{equation}
y_{\mathrm{D}}
=
r
+
\beta(1-\zeta)
\min_{i\in{1,2}}
Q_{\mathrm{C},i}
\left(
\mathbf{o}',\mathbf{d}^{+},\mathbf{p}^{+};
\bar{\psi}_{i}
\right),
\label{eq:cooperative_dra_target}
\end{equation}
where $\bar{\psi}_{i}$ denotes the parameters of the slowly updated
target critic. The DRA learns to value each AP–subcarrier decision together with its subsequent power allocation. The cooperative target combines the immediate reward with the CPA critics’ estimate of the complete next allocation, allowing discrete decisions to account for their consequences for power allocation.
Consequently, the DRA receives learning feedback that accounts for the
effect of its discrete decision on the subsequent power allocation and
the resulting long-term network reward.

The DRA branch loss is then given by:
\begin{equation}
\mathcal{L}_{\mathrm{D}}(\theta_{\mathrm{D}})
=
\mathbb{E}_{\mathcal{B}}
\left[
\frac{1}{|\mathcal{N}'|}
\sum_{n\in\mathcal{N}'}
\left(
Q_{\mathrm{D},n}
(\mathbf{o},d_n;\theta_{\mathrm{D}})
-
y_{\mathrm{D}}
\right)^{2}
\right].
\label{eq:cooperative_dra_loss}
\end{equation}
Therefore, an AP-subcarrier-count decision receives high value when
the subsequent CPA can use the selected configuration to obtain a high
long-term return. Conversely, the CPA evaluates the power
allocation in the context of the DRA configuration rather than
independently of the selected AP and subcarrier allocation. This
allows the 2 agents to learn complementary decisions while sharing
the same EE and QoS objective.

For comparison, the independent branching-DQN-SAC \cite{Tavakoli2018Branching,Haarnoja2018SAC} uses the
same state, action sequence, replay buffer, and common reward, but
does not use the CPA critic to update the DRA. Its DRA target is:
\begin{equation}
y_{\mathrm{D}}^{\mathrm{ind}}
=
r
+
\beta(1-\zeta)
\frac{1}{|\mathcal{N}'|}
\sum_{n\in\mathcal{N}'}
Q_{\mathrm{D},n}^{-}
\left(
\mathbf{o}',d_n^{+}
\right),
\label{eq:independent_dra_target}
\end{equation}
where $Q_{\mathrm{D},n}^{-}$ is the target branch network and
$\mathcal{N}'$ is the set of active users in the next state. The CPA still receives the executed DRA allocation since its power policy is conditioned on $\mathbf{d}$, but its critic feedback is not used in the DRA update. Thus, the cooperative formulation differs from the
independent one through the additional CPA critic feedback in \eqref{eq:cooperative_dra_target}.

The discrete feasibility layer and subcarrier mapper continue to
enforce C1-C3, while the power projection enforces C5-C6. The QoS
condition C4 is handled through the common reward in
\eqref{eq:mdp_reward}, since its satisfaction depends jointly on the
AP association, subcarrier allocation, transmit power, channel
conditions, and interference.

\subsection{Training and Online Operation}

DARA-DRL is trained centrally through interaction with the OWC environment. At each slot $t$, the observation $\mathbf{o}^{t}$ contains the measured state, and previous allocation. The DRA explores AP and subcarrier-count choices using $\epsilon$-greedy selection, followed by discrete feasibility processing and subcarrier mapping. Conditioned on the observation and resulting allocation $\mathbf{d}^{t}$, the CPA samples requested powers from its SAC policy and projects them onto the AP power budgets. The joint action $(\mathbf{d}^{t},\mathbf{p}^{t})$ is applied,
the common reward is calculated using \eqref{eq:mdp_reward}, and the reward and next observation complete the transition stored according to \eqref{eq:joint_transition}.

Both agents use the same replay-buffer mini-batch. The CPA updates its twin critics, actor and entropy temperature: critics use stored executed allocations, while the actor evaluates newly sampled, projected powers with entropy calculated before projection. The DRA uses the cooperative target in \eqref{eq:cooperative_dra_target}, where target CPA critics evaluate the complete next-state allocation. Target critics are soft-updated for stability. Training spans varying user loads, mobility, services and channels. Online, network parameters are fixed, the DRA acts greedily and the CPA produces deterministic powers. The same feasibility, subcarrier-mapping and power-projection layers enforce hard resource constraints. Allocations therefore require only forward inference and feasibility processing, without replay sampling, critic evaluation, learning updates or repeated solution of $\mathcal{P}_{1}$.

\subsection{Complexity Analysis}
For a single allocation interval, exhaustive
solution of the original mixed discrete-continuous problem
requires examining up to $(AL)^{N_{\max}}$ discrete allocations and globally solving the power-allocation subproblem for each. Its complexity is therefore
$\mathcal{O}((AL)^{N_{\max}}\mathcal{C}_{\mathrm{opt}})$,
where $\mathcal{C}_{\mathrm{opt}}$ is the cost of one global power-allocation solve. Reactive greedy allocation reduces this cost by scanning
the discrete candidates once, requiring
$\mathcal{O}(\nu)$, where $\nu=N_{\max}AL$.
It avoids neural-network inference and iterative optimisation.
PDP-OPA instead incurs
$\mathcal{O}(I_{\mathrm{PA}}\mathcal{C}_{\mathrm{PA}})$
for its power-allocation stage, where $I_{\mathrm{PA}}$
is the iteration count and $\mathcal{C}_{\mathrm{PA}}$
is the cost per iteration.

Learning-based schemes replace repeated optimisation with
trained-network evaluations; only online allocation cost is
considered here. For a fully connected network, one forward
pass costs $\mathcal{O}(\sum_r d_{r-1}d_r)$, where $d_r$
is the width of layer $r$. DARA-DRL requires
$\mathcal{O}(\mathcal{C}_{\mathrm{D}}
+\mathcal{C}_{\mathrm{P}}+\nu\log\nu+AL)$,
where $\mathcal{C}_{\mathrm{D}}$ and
$\mathcal{C}_{\mathrm{P}}$ are the DRA and CPA inference
costs. The remaining terms account for candidate sorting
and subcarrier assignment, assuming subcarrier orderings
are available; any additional interference-ranking cost
must also be included. Independent DQN--SAC has the same online complexity order
when its networks and feasibility processing match those
of DARA-DRL; cooperation changes their training updates.
P-DQN requires
$\mathcal{O}(\mathcal{C}_{\mu}+\mathcal{C}_{Q}+D)$,
where $\mathcal{C}_{\mu}$ and $\mathcal{C}_{Q}$ denote
the total parameter-network and Q-network evaluation costs,
and $D$ is the number of discrete choices compared.
Explicit joint-action encoding gives $D=(AL)^{N_{\max}}$.
Two-agent HAPPO requires
$\mathcal{O}(\mathcal{C}_{\pi_1}+\mathcal{C}_{\pi_2})$
for its two actor evaluations. Both schemes additionally
require feasibility processing. DARA-DRL therefore avoids exhaustive online search.
Although it incurs greater computational cost than the
greedy rule, it learns the long-term effects of current
allocations, including QoS shortfalls and allocation-change
costs, without requiring forecasts. Execution times are
compared in the results section.

\begin{table*}[t]
\caption{Simulation parameters}
\label{tab:simulation_parameters}
\centering
\scriptsize
\setlength{\tabcolsep}{4pt}
\renewcommand{\arraystretch}{1.15}
\begin{tabularx}{\textwidth}{
    @{}>{\raggedright\arraybackslash}X
       >{\raggedright\arraybackslash}X|
       >{\raggedright\arraybackslash}X
       >{\raggedright\arraybackslash}X@{}}
\hline
Parameter & Value & Parameter & Value \\
\hline
Episode length
& $1200$ slots
& EE reward normalisation
& Mean greedy EE during training \\

User arrival rate
& $1.5$ users/min
& Discount factor
& $0.99$ \\

Mean user residence time
& $4\,\mathrm{min}$
& Replay-buffer capacity
& $10^{6}$ transitions \\

User walking speeds
& $\{1,1.4,2\}\,\mathrm{m/s}$
& Mini-batch size
& $256$ \\

Mean pause duration
& $30\,\mathrm{s}$
& Target-network update coefficient
& $5\times10^{-3}$ \\

Service-class probabilities
& Equal: $1/3$ per class
& DQN learning rate
& $10^{-4}$ \\

SAC actor and critic learning rates
& $3\times10^{-4}$
& Warm-up duration
& $10^{4}$ steps \\

DQN exploration probability
& $1\rightarrow0.05$ over $2\times10^{5}$ steps
& SAC target entropy
& Negative of the active-user count \\

Training duration
& $10^{6}$ steps
& HAPPO update epochs
& $10$ \\

HAPPO clipping parameter
& $0.2$
& HAPPO advantage-estimation parameter
& $0.95$ \\
\hline
\end{tabularx}
\end{table*}

\section{Numerical Results and Discussion}
\label{sec:numerical_results}

\subsection{Simulation Setup}

This section evaluates the performance of DARA-DRL and whether it improves the long-term EE of the dynamic OWC network while satisfying user QoS requirements. DARA-DRL is compared with the 6 benchmarks defined in Sections~\ref{sec:optimisation_problem} and~\ref{sec:farad_rl}: the optimal solution, PDP-OPA, reactive greedy allocation, P-DQN, HAPPO and independent branching-DQN-SAC.  The room dimensions, AP layout, VCSEL array, ADR, Gaussian-beam channel, receiver-noise parameters, the receiver responsivity, FOV, modulation bandwidth, and remaining optical parameters are from \cite{Ncube2024Cells,Ncube2025Predictive} and the rest of the parameters are listed in Table~\ref{tab:simulation_parameters}. The DCO-OFDM size is $M=64$. The slot duration is $0.1\,\mathrm{s}$. User arrivals are Poisson and residence times are exponential, as in \cite{Ncube2025Predictive}. The 3 representative service classes chosen are video, general browsing and voice with a  minimum-rate vector of $[100,20,2]~\mathrm{Mbit/s}$, \cite{ITUT2021DigitalTV}. The DRA and CPA use two hidden layers of 256 ReLU units. The replay size, mini-batch size, SAC learning rate, discount factor and target-smoothing coefficient follow  \cite{Tavakoli2018Branching} and \cite{Haarnoja2019SAC}. Every learning based scheme is trained with 5 independent random runs, and the reported test results are averaged over 200 previously unseen episodes per run. The main performance metrics are long-term effective EE, QoS-satisfaction, and online decision time.

\subsection{Results}

\begin{figure}[t]
\centering 
\includegraphics[width=1\columnwidth,keepaspectratio]{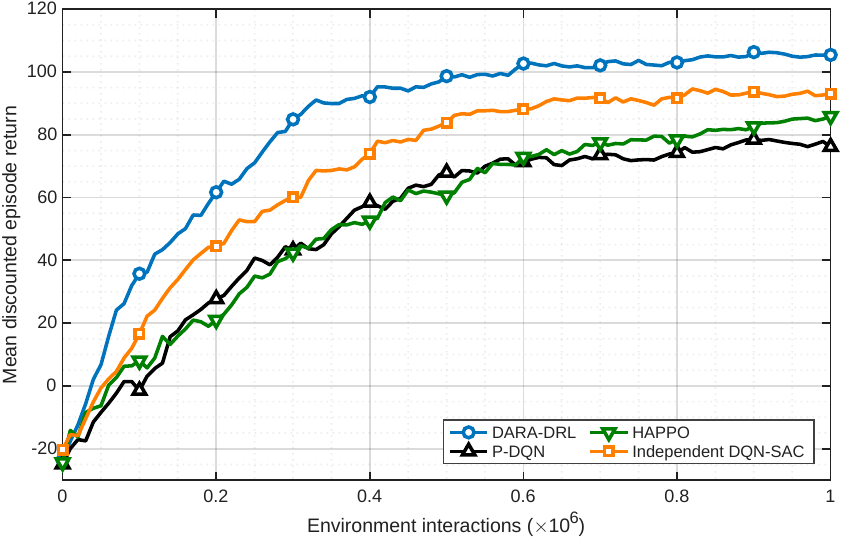}
\caption{DARA-DRL Training Performance.}
\label{fig:training_convergence} 
\vspace{-5mm}
\end{figure}

Figure~\ref{fig:training_convergence} shows the mean training return of the learning-based schemes over 5 runs. The horizontal axis is the number of interactions with the OWC environment, where each interaction corresponds to 1 time slot in which the controller observes the network state, selects the power and resource allocation, receives a reward  before observing the updated state at the beginning of
the next interval. The vertical axis is the unitless discounted episode return, \(G=\sum_{t=1}^{J}\beta^{t-1} r^{t}\) where $J$ is the number of slots in an episode. It is evaluated at the start of each episode, with \(j=1\). A higher return represents better long-term EE, QoS satisfaction and allocation stability. 

The initially negative returns occur because the untrained policies frequently violate the users' minimum-rate requirements. As training continues, all methods improve and their curves eventually stabilise. DARA-DRL converges fastest, stabilising after approximately \(5\times10^{5}\) interactions and reaching a final return of about 105. Independent DQN-SAC, HAPPO and P-DQN reach approximately 94, 88 and 80, respectively. The gap between DARA-DRL and independent DQN-SAC indicates the benefit of cooperative rather than independent agent training. HAPPO converges more slowly because it is not tailored to the sequential discrete-continuous problem, while P-DQN shows the difficulty of learning all 3 decisions through one hybrid agent. Overall, the results demonstrate that DARA-DRL learns a stable policy.

\begin{figure}[t]
\centering 
\includegraphics[width=1\columnwidth,keepaspectratio]{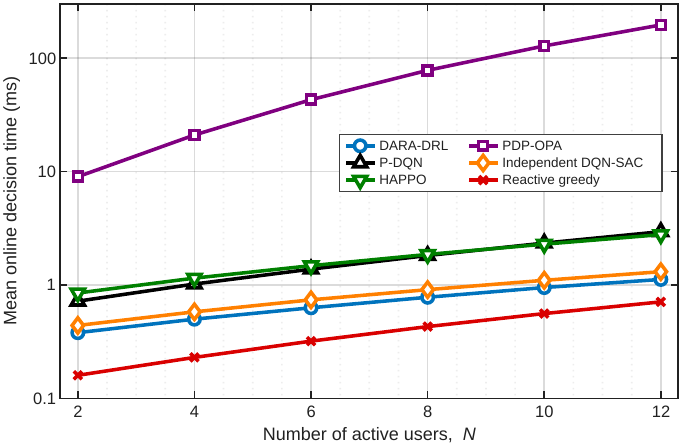}
\caption{Online computational performance.}
\label{fig:online_computational_performance} 
\vspace{-5mm}
\end{figure}

Figure~\ref{fig:online_computational_performance} shows how the online decision time changes with the number of active users. Reactive greedy is the fastest because it applies simple allocation rules without neural inference or iterative optimisation. DARA-DRL is the next fastest, increasing from \(0.38\) ms with 2 users to \(1.12\) ms with 12 users. Independent DQN-SAC has an almost similar runtime because it uses the same 2-agent structure. HAPPO and P-DQN follow with higher runtimes than Independent DQN-SAC
because their implemented policy-evaluation and action-selection steps
require more computation per decision. PDP-OPA increases much more rapidly, reaching \(196\) ms, because its iterative power-allocation procedure must be repeated whenever the network changes. DARA-DRL avoids repeated iterative optimisation through trained policy inference. Although reactive greedy is faster, DARA-DRL provides higher EE, QoS satisfaction and allocation stability in the evaluated scenarios.

\begin{figure}[t]
\centering
\subfloat[EE for general web browsing.]{
    \includegraphics[width=1\columnwidth,keepaspectratio]{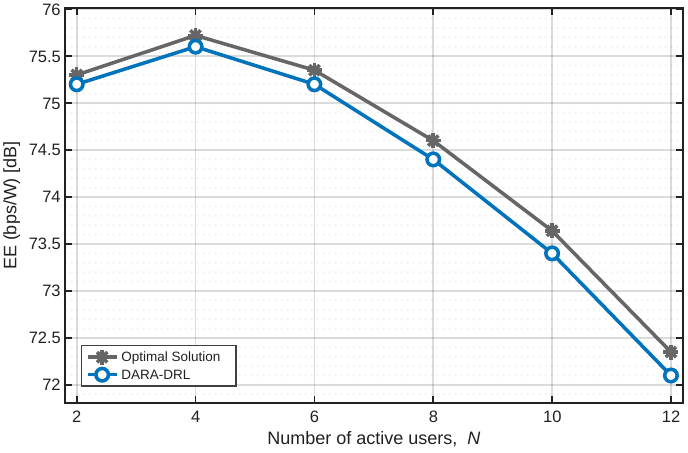}
    \label{fig:EEvsUsers_Opt}
}
\hfil
\subfloat[Gap from optimal solution]{
    \includegraphics[width=1\columnwidth,keepaspectratio]{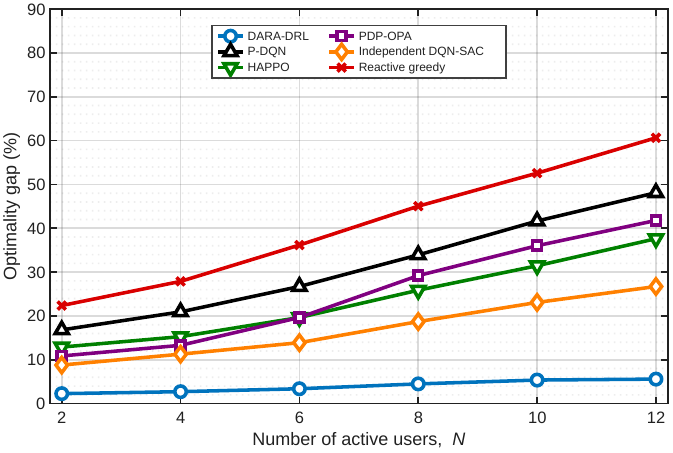}
    \label{fig:EE_gap_opt}
}
\caption{Gap from optimal solution}
\label{fig:small_system_validation}
\vspace{-5mm}
\end{figure}

Figure~\ref{fig:small_system_validation} compares the gap from
the optimal solution for 12 web users for the different schemes. In
Figure~\ref{fig:small_system_validation}(a), DARA-DRL closely follows the
optimal solution, with its optimality gap increasing only from \(1.8\%\)
to \(5\%\) as users increase. In Figure~\ref{fig:small_system_validation}(b), DARA-DRL maintains
the smallest optimality gap. It is closest to the
reference Optimal Solution at 12 users, with a gap of
approximately $5.6\%$, compared with $26.7\%$ for independent
DQN--SAC, $37.6\%$ for HAPPO, $41.8\%$ for PDP-OPA,
$48.1\%$ for P-DQN and $60.6\%$ for reactive greedy. These results show that DARA-DRL remains
close to the optimal solution as users increase.

\begin{figure}[t]
\centering
\subfloat[Energy-efficiency retention]{
    \includegraphics[width=1\columnwidth,keepaspectratio]{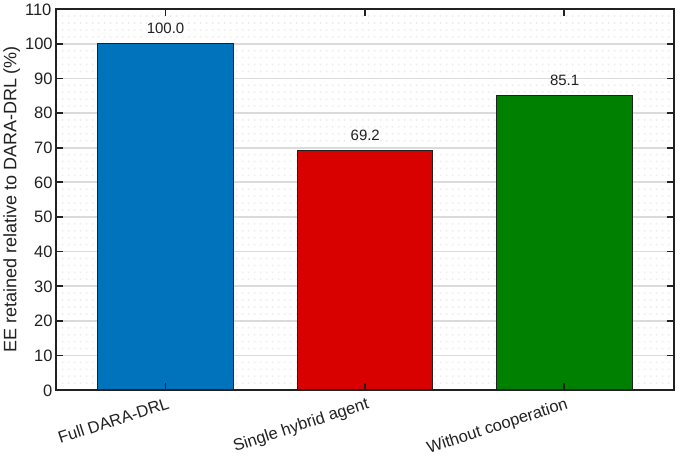}
    \label{fig:Energy-efficiency retention}
}
\hfil
\subfloat[AP reassignment frequency]{
    \includegraphics[width=1\columnwidth,keepaspectratio]{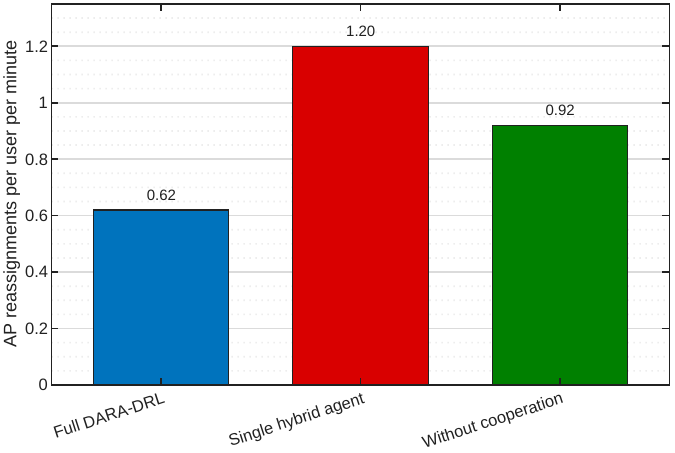}
    \label{fig:Handover_frequency}
}
\caption{Effect of agent specialisation and cooperative learning on DARA-DRL}
\label{fig:FARAD_RL_Cont}
\vspace{-5mm}
\end{figure}

To determine the contribution of each of the main DARA-DRL components, Fig.~\ref{fig:FARAD_RL_Cont} compares the full framework with 2 controlled configurations. The first replaces the 2 specialised agents with 1 hybrid agent, while the second retains the specialised agents but trains them independently. These configurations are evaluated under the same conditions. Long-term EE is measured over each \(120\)-s episode and averaged across the evaluation runs. It is then expressed relative to full DARA-DRL, while AP reassignments are measured as the average number of AP changes per user per minute.

Full DARA-DRL achieves the highest EE and the lowest AP reassignment rate of \(0.62\). Replacing the specialised agents with 1 hybrid agent gives the largest EE reduction, retaining \(69.2\%\), and produces \(1.20\) AP reassignments per user per minute. Training the 2 agents independently retains \(85.1\%\) of the EE and increases AP reassignments to \(0.92\). These results show that agent specialisation handles the discrete and continuous decisions more effectively, and cooperation ensures that the agents produce compatible decisions.

\begin{figure}[t]
\centering 
\includegraphics[width=1\columnwidth,keepaspectratio]{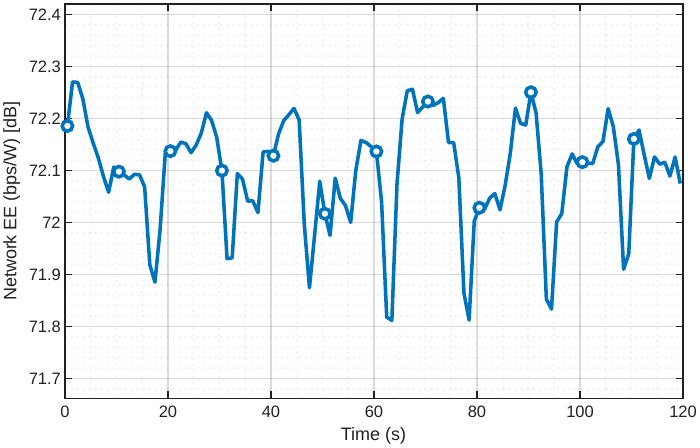}
\caption{Network energy efficiency of DARA-DRL over time under mixed user mobility.}
\label{fig:ee_over_time} 
\vspace{-5mm}
\end{figure}

Figure~\ref{fig:ee_over_time} shows the network EE achieved by DARA-DRL
over \(120\)~s with 12 active users; 4 are stationary
whilst 8 users are travelling at different speeds. Each
point is the EE averaged over 1 second. The EE fluctuates because user movement changes the optical channels,
interference and resulting allocation. The short reductions correspond to
temporary weak-channel conditions and resource adjustments, after which
DARA-DRL restores the EE. The lack of continuous degradation shows
that DARA-DRL maintains stable EE operation under mixed
user mobility.

\begin{figure}[!t]
\centering
\subfloat[Network energy efficiency.]{
    \includegraphics[width=1\columnwidth,keepaspectratio]{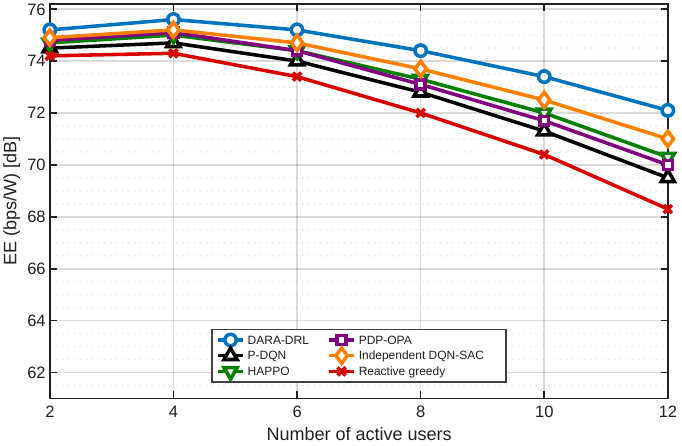}
    \label{fig:EEvsUsers}
}
\hfil
\subfloat[QoS satisfaction]{
    \includegraphics[width=1\columnwidth,keepaspectratio]{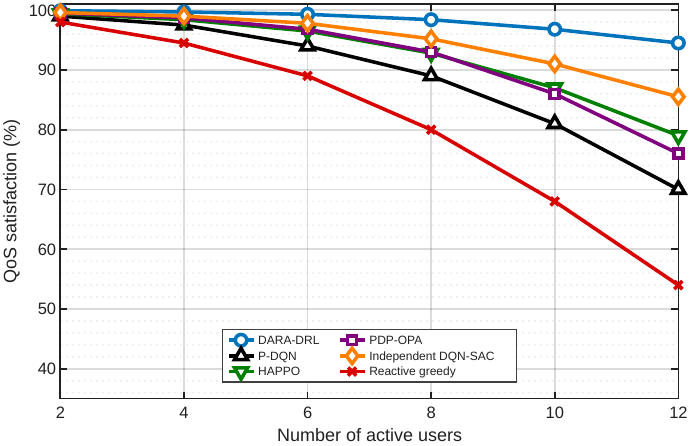}
    \label{fig:QoS}
}
\caption{EE Performance under increasing web-user load}
\label{fig:ee_under_varying_load}
\vspace{-5mm}
\end{figure}

Figure~\ref{fig:ee_under_varying_load} compares EE and QoS satisfaction
for web users. From Figure~\ref{fig:ee_under_varying_load}(a), EE increases between 2 and 4 users as more data is delivered
for little additional power. Above 4 users, power consumption grows
faster than the delivered data rate, reducing EE. DARA-DRL maintains the
highest EE relative to other schemes. In Figure~\ref{fig:ee_under_varying_load}(b) the QoS satisfaction rate is:
$\mathrm{QoS\ satisfaction}=
\frac{\text{number of users meeting the minimum rate}} {\text{total number of active users}}\times100\%$, and averaged over evaluation slots and runs. DARA-DRL achieves higher QoS satisfaction than the benchmarks as its cooperating agents jointly adapt AP, subcarrier and power decisions using current measurements and previous allocations.

\section{Conclusion}

This work investigated energy-efficient AP association, subcarrier allocation and power control in a dynamic OWC network with mobile users who subscribe to different service classes. The resulting mixed discrete-continuous and time-dependent problem was addressed using DARA-DRL, in which a discrete resource-allocation agent cooperates with a continuous power-allocation agent. Feasibility
layers enforce the association, subcarrier and AP power
constraints, while the common reward penalises violations
of the users' minimum-rate requirements. The results show that DARA-DRL approaches the optimal solution and provides higher EE, QoS satisfaction than state-of-the-art learning-based and conventional benchmarks. It also produces allocations within a few milliseconds, avoiding the repeated iterative optimisation required by conventional methods. Two main lessons can be drawn. First, using specialised agents is more effective than forcing one agent to learn different action types. Second, the agents must be trained cooperatively because the value of an association or subcarrier decision depends on the power subsequently allocated. Future work will consider experimental validation and also be extended to consider optimising power on individual subcarriers and coordinating subcarrier reuse across APs for improved interference management and EE.

\bibliographystyle{IEEEtran}
\bibliography{Learning}

\end{document}